\documentclass{ws-tpe}
\usepackage{multicol}
\begin{document}

\catchline{1}{1}{2019}{}{}
\markboth{Overduin et al}{The Scourge of Online Solutions...}

\title{The Scourge of Online Solutions and\\an Academic Hertzsprung-Russell Diagram}

\author{James Overduin$^{\ast}$, 
Jacob Buchman, Jonathan Perry and Thomas Krause}

\address{Department of Physics, Astronomy and Geosciences, Towson University\\
8000 York Road, Towson, Maryland 21252, USA\\
$^{\ast}$\email{joverduin@towson.edu}}

\maketitle


\begin{abstract}
We report on preliminary results of a statistical study of student performance in more than a decade of calculus-based introductory physics courses.
Treating average homework and test grades as proxies for student effort and comprehension respectively, we plot comprehension versus effort in an academic version of the astronomical Hertzsprung-Russell diagram (which plots stellar luminosity versus temperature).
We study the evolution of this diagram with time, finding that the ``academic main sequence'' has begun to break down in recent years as student achievement on tests has become decoupled from homework grades.
We present evidence that this breakdown is likely related to the emergence of easily accessible online solutions to most textbook problems, and discuss possible responses and strategies for maintaining and enhancing student learning in the online era.
\end{abstract}


\begin{multicols}{2}
\section{Introduction}

Two of us have been teaching introductory calculus-based physics courses at Towson University for a combined 55 years.
Like most physics teachers, we require students to turn in regular homework assignments for a significant portion of their overall course grade.
Some of the problems in these assignments are our own, while others are drawn from textbooks.
We aspire to regenerate these problem sets from scratch for each course, every semester; but given the realities of teaching workloads at a comprehensive four-year university, the replacement rate is typically less than 100\%.

The purpose of homework is simple: there is no other way for students to master the material than by spending significant time working through problems on their own, especially in a subject as deductive as this one.
The amount of course credit attached to these homework assignments should be high enough to motivate students to put in the necessary time.
But it should not be high enough to play a decisive role in determining a student's final grade, since there is no way for instructors to be certain that students have done the work themselves.
In practice, we have found over the years that best results are achieved when homework counts for about 20\% of a student's overall course grade.

What do we mean by ``best''?
Like most teachers, we rely on a mixture of clues, including student feedback as well as scores on tests and quizzes.
We pause at intervals during lectures and check class comprehension using interactive multiple-choice concept (``clicker'') questions.
We use standardized assessments such as the Force Concept Inventory (FCI)\cite{fci} and the Conceptual Survey in Electricity and Magnetism (CSEM),\cite{csem} comparing student responses before and after the course.
And we study grade distributions carefully, checking whether they conform to our expectation of a close-to-normal distribution.

In recent years, one of us has noticed a disturbing trend in these grade distributions.
What was once a single, close-to-normal distribution (with a peak typically somewhere in the 70\% range) has gradually bifurcated into a bimodal distribution, with one peak closer to the 80\% range and the other typically in the 50\%-60\% range.
Informal investigation suggests that students in the first group are those who spend significant time on homework, while those in the second either turn in very little homework, or have access to solutions which they copy without learning how to solve the problems themselves.
To use a time-worn analogy, these students are like aspiring athletes who hope to achieve success by going to the gym and watching others do the exercises for them.

One possible culprit is the rise of easily accessible solutions to most textbook (and many instructor-created) problems on internet websites such as \url{chegg.com} and \url{coursehero.com} (Fig.~\ref{fig-websites}).
\begin{figurehere}
\centerline{
\includegraphics[width=2.2in]{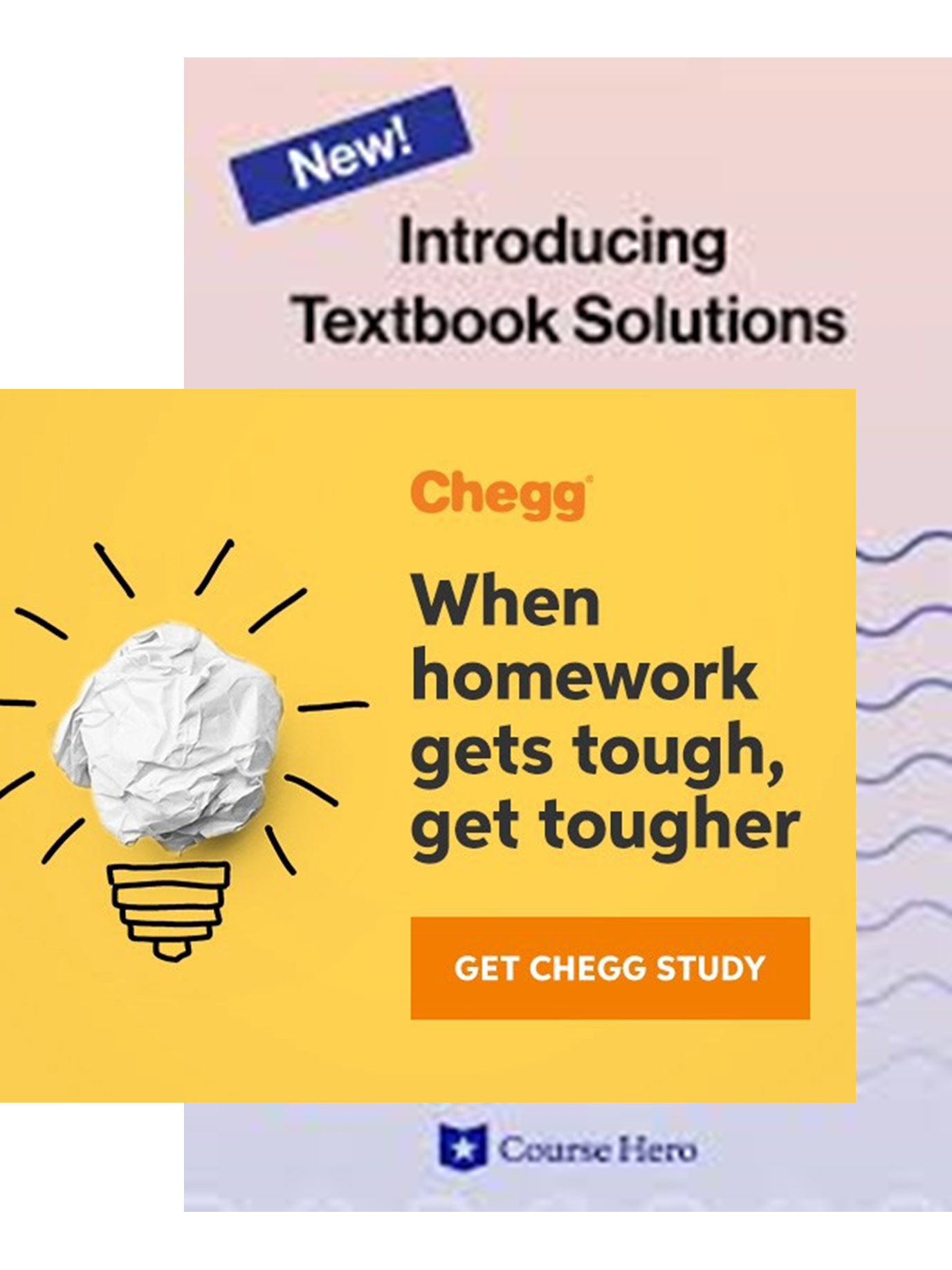}
}
\caption{Images from online solutions websites {\tt chegg.com} and {\tt coursehero.com}}
\label{fig-websites}
\end{figurehere}
\noindent These websites typically claim to help customers ``study smarter, not harder.''
They may indeed be of some help to disciplined students, who turn to them only after first making a genuine effort on their own.
But for other students (presumably those in the second ``hump'' of the bimodal distribution),  the temptation of a shortcut to the answer proves irresistible, robbing them of the essential ingredient to all successful learning: {\em the struggle}.
Students who miss this moment are like tourists who have been guided to their destination with blindfolds on.
If they are then dropped anywhere else in the city on their own, they are helpless, because they have not explored on their own.
All physics teachers are familiar with the student who complains, ``This test had nothing to do with the homework! In the homework, the train went east, but here it went west!''

But how to test such a hypothesis?
And what to do about it?
In this short paper, we mostly focus on the first question, although we consider some possible ideas in the concluding discussion.
We look forward to hearing about others from readers.

\section{A Tool from Astronomy}

To help us understand what is happening in these all-important introductory physics courses, we turned to a familiar tool from astronomy.
Developed between 1908 and 1913 by Hans Rosenberg,\cite{Rosenberg1910} Ejnar Hertzsprung,\cite{Hertzsprung1911} and Henry Norris Russell,\cite{Russell1913} the Hertzsprung-Russell (HR) diagram has become essential to the study of stars and their evolution.\cite{Nielsen1963,Sitterly1970,DeVorkin1978,Gingerich2013}
It plots stellar luminosity or magnitude (on the vertical) against temperature (or color, or spectral class) on the horizontal (Fig.~\ref{fig-originalHRdiagram}).
\begin{figurehere}
\centerline{
\includegraphics[width=3.3in]{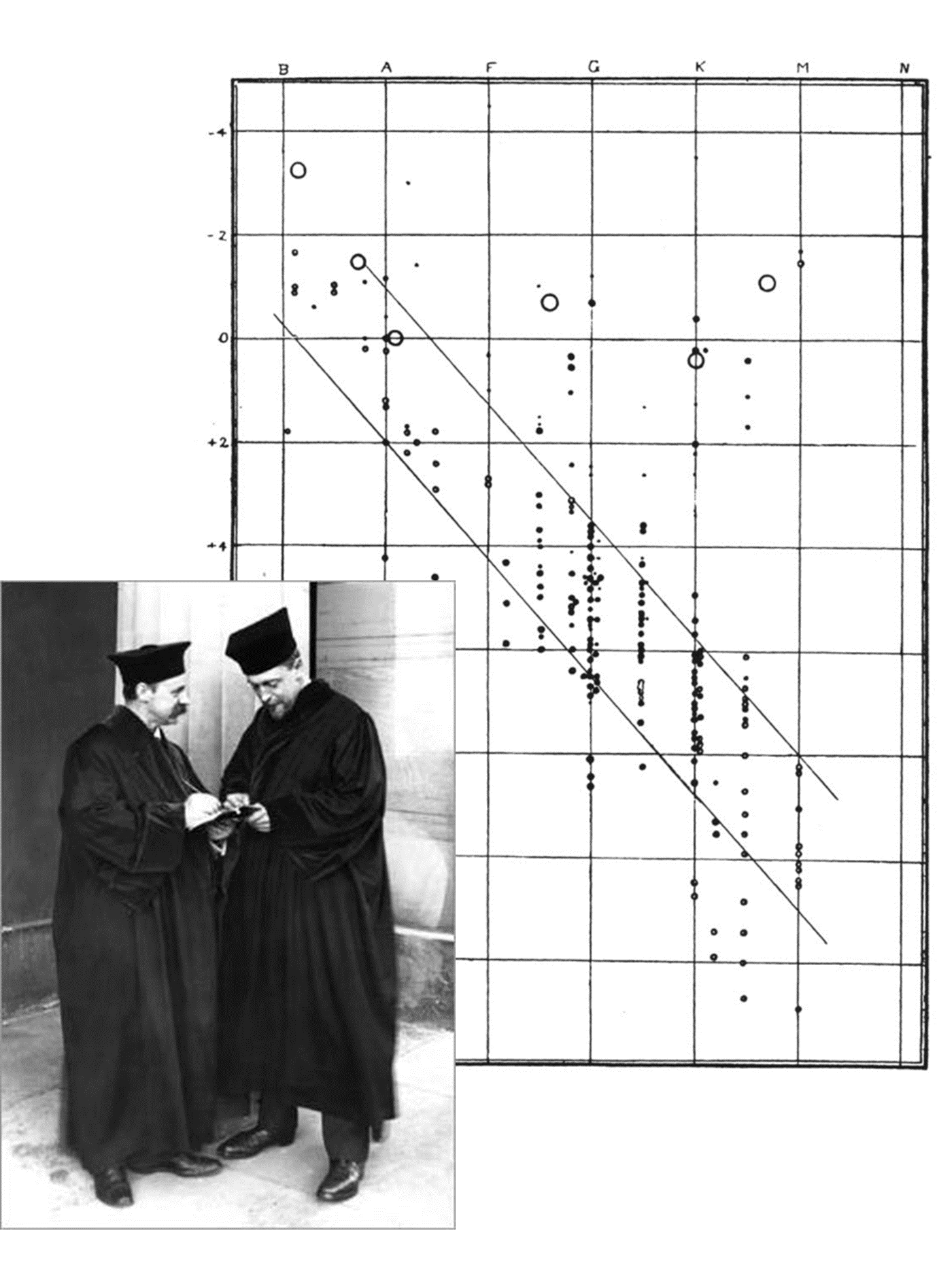}
}
\caption{Early HR diagram by Russell plotting apparent magnitude versus spectral class. The inset shows Hertzsprung in 1909 (right) with Karl Schwarzschild, later the discoverer of black holes in general relativity, who played a critical role in introducing Russell to Hertzsprung's work.}
\label{fig-originalHRdiagram}
\end{figurehere}

Generally speaking, the hotter a star is, the greater its luminosity.
Most stars thus comprise what is known as the ``main sequence,'' an approximately linear band running diagonally across the HR diagram.
(For historical reasons, stellar temperatures are ordered from hot to cool going from left to right, so that the slope of the main sequence looks negative on the diagram, but is really positive in the sense of luminosity versus temperature.)
There are important exceptions, chiefly the giants (stars that are luminous despite their cool temperatures) and dwarfs (stars that are faint despite their hot temperatures).
Modern HR diagrams use data from orbiting telescopes like the {\it Gaia} satellite\cite{Gaia} to plot magnitudes and colors for millions of stars within a thousand light-years of the Sun (Fig.~\ref{fig-gaiaHRdiagram}).

\begin{figurehere}
\centerline{
\includegraphics[width=3.1in]{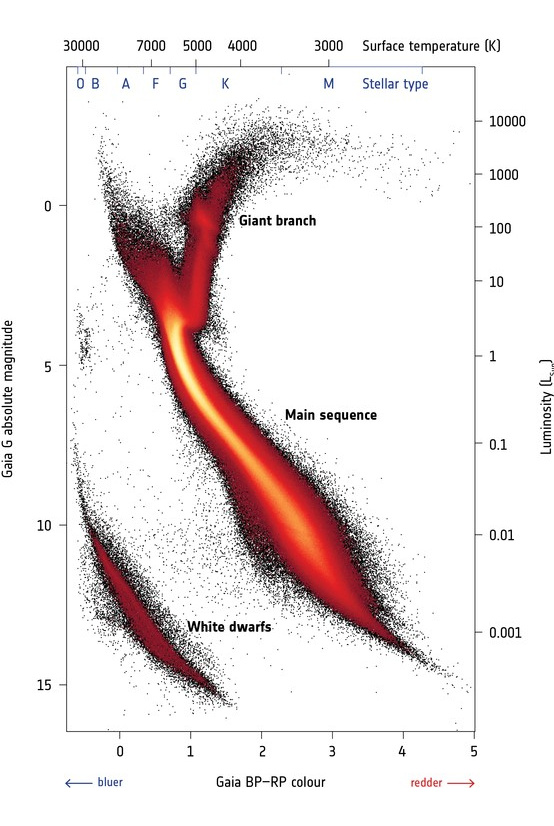}
}
\caption{Modern version of the HR diagram, in which each of the over 4~million pixels corresponds to a real star observed by ESA's {\it Gaia} satellite.}
\label{fig-gaiaHRdiagram}
\end{figurehere}

Although the physics of human behavior is likely more complex than that governing spherical balls of hydrogen and helium, it occurred to us that student learning might be studied in the same way.
The natural analog of stellar temperature would be student {\em effort}, as measured (at least in an ideal world) by average homework grade.
And the corresponding analog of stellar luminosity would be student {\em achievement}, as measured by average grade on tests and exams.
Plotting one against the other, we wondered if we would find an ``academic main sequence'' analogous to the stellar one?
And if so, could it be used to study the evolution of student learning in a changing academic landscape?

\section{The Academic HR Diagram}

Our data comprises twelve semesters of calculus-based Introductory Physics 1 and 2 courses (Mechanics and Electricity and Magnetism respectively) from 2012-2020 at Towson University, which has been among the ten largest undergraduate programs among primarily Masters-granting institutions in the U.S.A. throughout this period.\cite{apsStatistics}
Altogether, the dataset incorporates anonymized average homework and test grades for 364 Physics~1 students and 261 Physics~2 students.

Academic HR diagrams for all these students are shown in Fig.~\ref{fig-academicHRdiagrams}, with best-fit linear ``main sequences'' superimposed on the data (obtained from straightforward least-squares regression).
\begin{figurehere}
\centerline{
\includegraphics[width=3.4in]{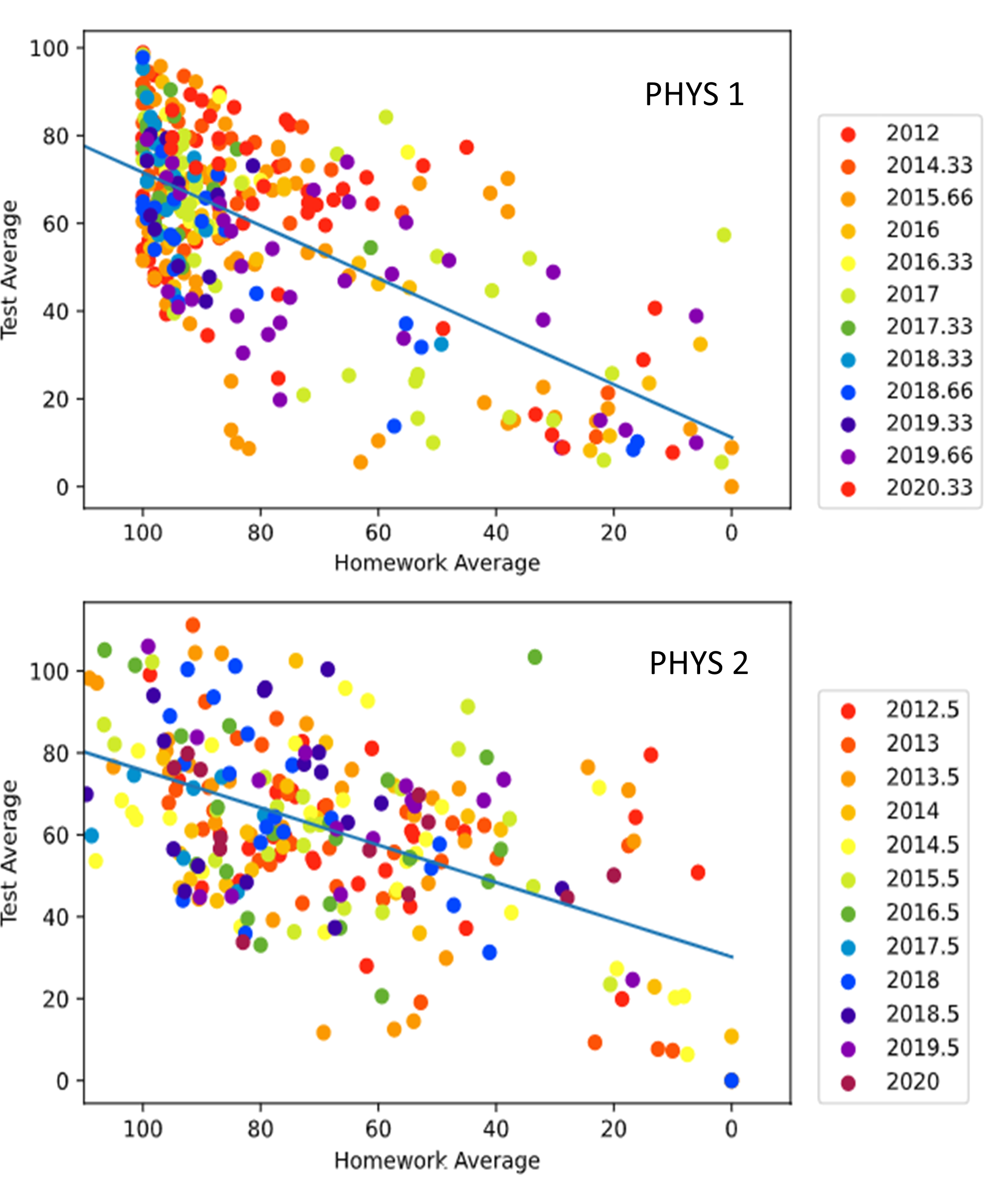}
}
\caption{Aggregate academic HR diagrams for calculus-based Introductory Physics students at Towson University}
\label{fig-academicHRdiagrams}
\end{figurehere}
\noindent As expected, the academic HR diagram shows significantly more scatter than the astronomical one.
Nevertheless, it is clear that, on average, more effort on homework (student ``temperature'') does correlate with higher test scores (``luminosity'').
There is some evidence, perhaps marginal given the sample size, for populations corresponding to astronomical ``giants'' (i.e., students who excel on tests despite expending little effort on homework) and ``dwarfs'' (those whose test scores suggest that they have not learned the material, despite having apparently exerted great effort on homework).
These trends persist across teaching styles, as shown in the diagrams by the fact that the PHYS~1 instructor adopts an upper limit of 100\% for both homework and test grades, while the PHYS~2 instructor favors the use of bonus problems so that either grade can sometimes exceed 100\%.

Some differences between the academic HR diagrams in Fig.~\ref{fig-academicHRdiagrams} are likely attributable to adaptive behavior by students.
Some students at the lower end of the main sequence in PHYS~1, for instance, may conclude that they should spend more time on homework, and climb up toward the middle in PHYS~2.
Similarly, some students at the upper end of the main sequence in PHYS~1 may conclude that they can afford to spend less time, and slide down toward the middle in PHYS~2.
Students in the dwarf group in PHYS~1 may be motivated to do their homework more effectively, and thus migrate vertically upward and join the main sequence---or they may conclude it is not worth the effort and drop out.
Either way, we would expect the dwarf group to be smaller for PHYS~2 than for PHYS~1, and for the PHYS~2 diagram to be more ``bunched up'' toward the middle than the PHYS~1 diagram.
All these features are evident in Fig.~\ref{fig-academicHRdiagrams}.

We have color-coded the datapoints in Fig.~\ref{fig-academicHRdiagrams} by semester, so that one can in principle follow changes in the academic HR diagram with time.
When the diagram is animated, so that individual semesters are superimposed on top of each other in sequential order, scatter is seen to grow steadily larger with time.
This is hard to convey in a static image, but suggests that the correlation between homework grade (as an indicator of student effort) and comprehension is weakening with time.
A natural hypothesis to consider is that homework grades are no longer an accurate reflection of actual effort, as could happen if students increasingly have access to fully worked-out solutions.

To test this hypothesis, we begin by noting that our assumed main sequence has a simple linear form, $y=mx+b$, where $y$ is test grade (``comprehension''), $m$ is slope, $x$ is homework grade (``effort''), and $b$ is the value of the $y$-intercept.
For reasons to be explored further below, it is convenient to break the fits into two periods: before and after 2016.
Average best-fit values of $m$ and $b$ for both periods are tabulated in Table~1, where $\Delta$ is the relative standard error, which we take here as a measure of uncertainty or ``badness of fit.''

\begin{tablehere}
\tbl{Best-fit main sequence parameters}
{\begin{tabular}{@{}llccc@{}}
    \toprule & & $m$ & $b \mbox{ (\%)}$ & $\Delta \mbox{ (\%)}$ \\
    \colrule \mbox{PHYS 1} & \mbox{(before 2016)} & 0.55 & 16 & 17 \\
    & \mbox{(after 2016)} & 0.60 & 13 & 40 \\
    \colrule \mbox{PHYS 2} & \mbox{(before 2016)} & 0.44 & 30 & 30 \\
    & \mbox{(after 2016)} & 0.34 & 40 & 74 \\
    \botrule
    \end{tabular}}
\end{tablehere}

Table~1 suggests an inverse correlation between the values of $m$ and $b$, as might be expected.
Thus, PHYS~1 shows a steeper slope and correspondingly smaller intercept than PHYS~2.
This difference is likely a reflection of differing teaching styles, in particular the more liberal use of bonus problems in homework assignments than on tests (thus lowering the slope in PHYS~2 relative to PHYS~1).
The correlation is confirmed in Fig.~\ref{fig-slopeVsIntercept}, which plots best-fit slope versus intercept in each semester for both courses.
The size of each datapoint in this and subsequent plots is scaled to the number of students in the class per semester.
Fig.~\ref{fig-slopeVsIntercept} establishes that the academic main sequence can be effectively characterized by a single degree of freedom ($m$ or $b$).
In what follows, we express our results in terms of the slope $m$.

\begin{figurehere}
\centerline{
\includegraphics[width=2.9in]{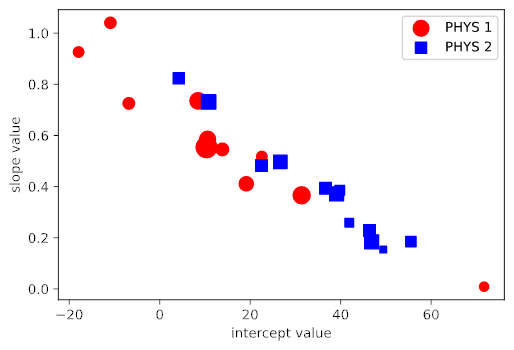}
}
\caption{Correlation of best-fit slope and intercept for the academic main sequences in Fig.~\ref{fig-academicHRdiagrams}.}
\label{fig-slopeVsIntercept}
\end{figurehere}

\section{Impact of Online Solutions}

The most striking aspect of the numbers in Table~1 is the tremendous increase in relative standard error $\Delta$ with time.
(Thanks to the correlation established above, this quantity has the same value whether standard error is taken relative to slope or intercept.)
Over the period 2016-2020, the ``badness of fit'' of the academic main sequence jumped from 17\% to 40\% for Physics~1, and from 30\% to 74\% for Physics~2, relative to its value from 2012-2016.

To follow this change more closely in time, we show in Fig.~\ref{fig-slopeVsTime} a plot of best-fit slope $m$ for each semester, with the size of each errorbar given by $\Delta$. This figure confirms that until about 2016, the academic main sequence is reasonably well fit by a simple linear function with consistent values of $m$ and $b$.
That is, until that time, an average student's effort on homework did serve as a faithful predictor of their subsequent achievement on tests and exams.
After that time, however, best-fit values of $m$ and $b$ began to fluctuate, and their relative uncertainties grew tremendously.
Indeed, there are some individual courses and semesters where the ``main sequence'' resembles a vertical line, so that there is no functional relationship between homework and test grades at all (e.g., where nearly all the students scored homework averages near 90\%, yet their test grades varied between 40\% and 100\%).

\begin{figurehere}
\centerline{
\includegraphics[width=3.4in]{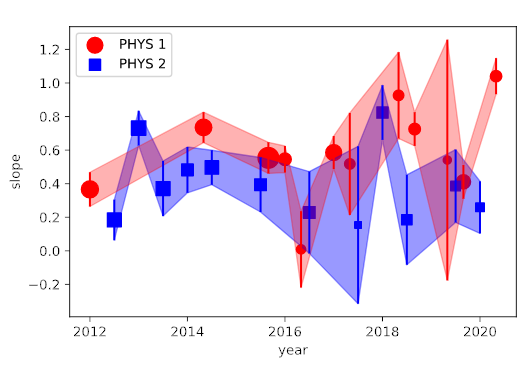}
}
\caption{Best-fit slope $m$ of the academic main sequence as a function of time, with the size of the error bars given by the ``badness of fit'' (relative standard error) $\Delta$.}
\label{fig-slopeVsTime}
\end{figurehere}

To test the hypothesis that this breakdown is related to the emergence of online solutions, we turned to the website \url{SEMrush.com}, which tracks internet traffic to any given address.
We obtained historical data on traffic to the two leading internet solutions vendors, \url{coursehero.com} and \url{chegg.com}, and digitized it.
Fig.~\ref{fig-scatterVsWebTraffic} plots the results and superimposes them on our datapoints for standard error.
The overlap is remarkable, particularly during the years 2016-2018 when the use of online solutions appears to have become truly widespread.
We regard this as an interesting, if preliminary result that bears further investigation, and we invite other instructors to see if they can duplicate and/or expand on our findings.

\begin{figurehere}
\centerline{
\includegraphics[width=3.4in]{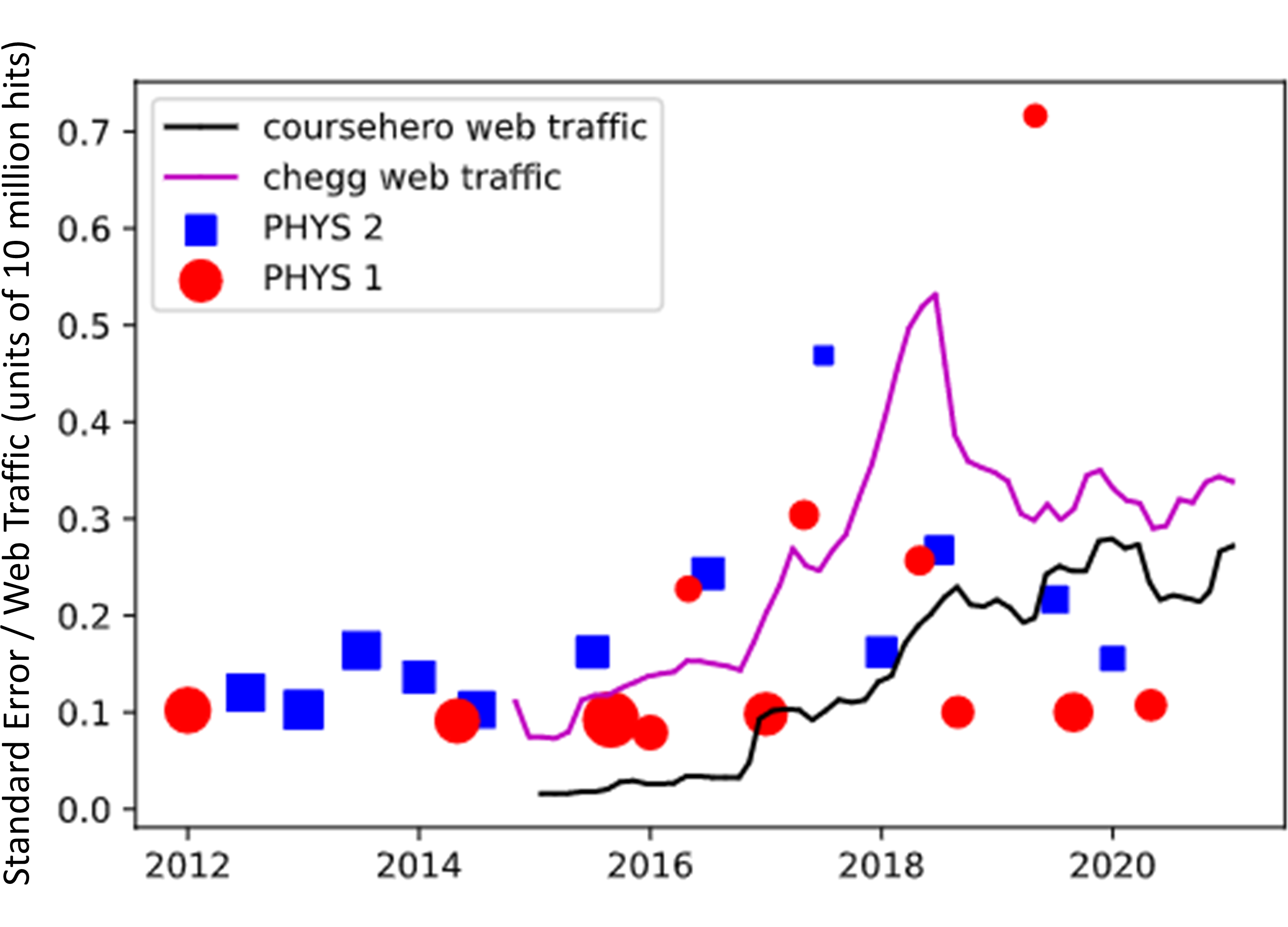}
}
\caption{Standard error for our academic HR diagrams (points), superimposed on a plot of web traffic to the two leading internet solutions providers (solid lines, measured in units of 10 million hits).}
\label{fig-scatterVsWebTraffic}
\end{figurehere}

\section{Discussion}

There is good evidence for an academic Hertzsprung-Russell diagram, with homework grade serving as a proxy for student effort and mean test grade measuring student comprehension.
But the correlation (i.e., the existence of a stable linear ``main sequence'' incorporating most students) has started to break down.
The growing ``badness of fit'' (standard error) of this main sequence appears to be correlated with increased web traffic to the most popular online solution vendors (\url{chegg.com} and \url{coursehero.com}), particularly since 2016.
If so, then the traditional homework model may no longer be an effective way to promote student learning for most students.

What are the practical solutions to this problem, if any?
Ideally, of course, instructors could respond by generating entirely new homework problem sets from scratch, for every course, every semester, so that online vendors would always be a step behind.
This is unrealistic for instructors at most institutions, who must teach multiple courses at once.
Another approach might be to reduce the fraction of the overall course grade associated with homework (say from 20\% to 10\%, or even nothing; hoping that students will understand that doing the homework will be essential to performing well on tests).
This, however, effectively shifts the weight of a student's overall course grade further toward tests and exams, which would be counter-productive for most introductory students.

A third idea, and one that we have begun to adopt in practice, is to supplement or replace the homework portion of our courses with in-class assessments like low-stakes quizzes or group problem-solving exercises at the whiteboard.
There are some clear benefits to this approach.
However, there are downsides as well: the kinds of problems that can be worked by students this way are necessarily shorter and often more conceptual.
Much of the physical insight that we want our students to discover comes through longer, more context-rich problems whose solution requires more time.
And in general, any additional in-class exercise necessarily reduces the amount of material that the instructor can cover over the duration of the course.

It may be that there is no ideal solution, and that we must learn to take comfort in the words of legendary physicist and teacher Victor Weisskopf: ``Better to uncover a little, than to cover a lot.''
We look forward to hearing from others about ways that they have adapted their introductory physics courses to the online era.

\section*{Acknowledgments}

We thank the Maryland Space Grant Consortium and Fisher College of Science and Mathematics at Towson University for support.

\end{multicols}

\begin{thebibliography}{10}
\bibitem{fci} D. Hestenes, M. Wells and G. Swackhammer, ``Force concept inventory,'' {\it The Physics Teacher} 30 (1992) 141-158
\bibitem{csem} D.P. Maloney, T.L. O'Kuma, C.J. Hieggelke and A. Van Heuvelen, ``Surveying students' conceptual knowledge of electricity and magnetism,'' {\it Physics Education Research, American Journal of Physics Supplement} 69 (2001) S12-S23
\bibitem{Rosenberg1910} H. Rosenberg, ``\"Uber den Zusammenhang von Helligkeit und Spektraltypus in den Plejaden'' (``On the relationship between brightness and spectral type in the Pleiades''), {\it Astronomische Nachrichten} 186 (1910) 71-77 [see translation by J. Hollan at \url{https://www.leosondra.cz/en/first-hr-diagram/} (accessed June 1, 2021)]
\bibitem{Hertzsprung1911} E. Hertzsprung, ``\"Uber die Verwendung Photographischer Effektiver Wellenlaengen zur Bestimmung von Farbenaequivalenten'' (``On the use of photographic effective wavelengths for the determination of color equivalents''), {\it Publikationen des Astrophysikalischen Observatoriums zu Potsdam} 22 (1911) 63
\bibitem{Russell1913} H.N. Russell, ```Giant' and `dwarf' stars,'' {\it The Observatory} 36 (1913) 324-329 [This article reports on Russell's address to the Royal Astronomical Society on June 13, 1913, but does not reproduce his slides containing the first HR diagram. That diagram is however reproduced in the review by Nielsen.\cite{Nielsen1963}]
\bibitem{Nielsen1963} A.V. Nielsen, ``Contributions to the history of the Hertzsprung-Russell diagram,'' {\it Centaurus} 9 (1964) 219-253
\bibitem{Sitterly1970} B.W. Sitterly, ``Changing interpretations of the Hertzsprung-Russell diagram,'' {\it Vistas in Astronomy} 12 (1970) 357-366
\bibitem{DeVorkin1978} D.H. DeVorkin, ``Steps toward the Hertzsprung-Russell diagram,'' {\it Physics Today} 31 (1978) 32-39
\bibitem{Gingerich2013} O. Gingerich, ``The critical importance of Russell's diagram,'' in M.J.~Way and D.~Hunter (eds.), {\it Origins of the Expanding Universe: 1911-1932}, ASP Conference Series, Vol.~471 (San Francisco: Astronomical Society of the Pacific, 2013); arXiv:1302.0862
\bibitem{Gaia} Gaia Data Processing and Analysis Consortium (DPAC); Carine Babusiaux, IPAG – Université Grenoble Alpes, GEPI – Observatoire de Paris, France; \url{https://sci.esa.int/web/gaia/-/60198-gaia-hertzsprung-russell-diagram} (accessed June 1, 2021)
\bibitem{apsStatistics} American Physical Society website, ``Top Educators,'' \url{https://www.aps.org/programs/education/statistics/topproducers.cfm} (accessed June 1, 2021)
\end{thebibliography}
\end{document}